\documentstyle{article}

\author{Christof Zalka \\ zalka@t6-serv.lanl.gov}

\title{Could Grover's quantum algorithm help in searching an 
actual database?}

\begin{document}
\maketitle
\begin{abstract}
I investigate whether it would technologically and economically make
sense to build database search engines based on Grover's quantum
search algorithm. The answer is not fully conclusive but in my
judgement rather negative.
\end{abstract}

\section{Introduction}
In the quantum computing community different views about the
usefulness of Grover's algorithm for actual database search have been
expressed informally but so far nothing has been published on the
subject. Probably most authors (e.g. \cite{grover1} \cite{smolin}) who
refer to Grover's quantum search algorithm as ``database search''
think of the oracle as a ``virtual database'' and don't mean to imply
actual database search.

The main argument against building a database search engine based on
Grover's algorithm is of course that the whole addressing system would
have to work in superposition, thus it would have to be built out of
(presumably expensive) quantum hardware which would have roughly the
size of the database. At this point one wonders whether a (partially)
parallel classical computer with many processors would not be a better
solution. Here I analize and compare in more detail the cost and
performance of quantum and classical search engines.

\section{Grover's algorithm}

Grover's algorithm \cite{grover1,zalka} a priori solves a somewhat
artificial problem: We are given some quantum hardware around which
there is is a black box so we don't know how it works. The hardware
outputs a 0 or a 1 for any of a large number of possible inputs. The
task is to find an input for which the output is 1. If the number
$n_1$ of such inputs is known, Grover's algorithm finds one of them
querying the black box (also called ``oracle'') some $\pi/4
\sqrt{N/n_1}$ times. Classically the number of queries would on
average be more like $N/n_1$.

\subsection{what is it good for?}

Practically there are computational problems where we don't know
better that to try through a large number of possibilities to find one
that suits us. On the other hand often (possibly heuristic) classical
algorithms are used which give a much better than square root
improvement over a simple ``exhaustive'' search.

I have been told about the following examples of problems where no
(much) better algorithm than a simple unstructured search is known and
where thus Grover's algorithm could help:

\begin{itemize}

\item{breaking a DES (data encryption standard) key}

\item{certain (particularly hard) instances of the SAT
(satisfyability of boolean formulae) problem}

\item{hard instances of coloring a map with just 4 colors, 
a task which is guaranteed to be possible by the 4-color theorem}

\end{itemize}

\section{Comparing classical and quantum search engines}

\subsection{preliminary considerations}

An existing database (e.g. everything available via the internet) of
course can't be searched with Grover's algorithm as the addressing
system (the internet) is not quantum. Thus here I only consider
purpose built databases including an addressing system and search
engine. 

Also we only consider unordered databases as clearly an ordered
database can be searched in logarithmic (in the number of datasets)
time.

Actually I only consider databases which \underline{can't} be ordered
because ordering a databases once and for all is not such a big task
and can actually be done while the datasets are acquired at a cost of
$\log N$ per dataset where $N$ is the size of the database.

A database can't be ordered when our search criteria are too
complex. Still, often indexing according to different search criteria
is possible which allows searching in logarithmic time for a not too
big variety of search criteria. Here I consider the case where our
search criteria are so complex and varied that this is not possible.

\subsection{definition of the problem}

We have a database consisting of $N$ datasets, each of size $d$ (say
$d$ bits). The search criteria are binary functions to be computed
from the datasets which tell us whether a dataset meets our criteria
or not. The size (e.g. number of gates) of a processor needed to
compute the search criterion is $p$ and the computation time is $t$. I
don't consider the possibility of space time tradeoffs between these 2
quantities.

\subsection{how to compare quantum and classical in a fair way}

I compare cost and performance of the best classical search engine I
can think of to the best quantum engine (based on Grover's algorithm)
I can think of. The database is of course the same in both cases. We
could then e.g. compare the performance (speed) of the quantum and
classical solutions for a given and equal cost of the machines. A
problem is that we don't know the (probably high) cost of future
quantum hardware.

Alternatively we could compare for a given performance (= search
speed). Again a problem is that we don't know how the clock rate of
quantum hardware will compare to that of classical hardware. Actually
from the various quantum hardware proposals one get's the impression
that it might be rather slow due to the slowness of 2-qubit gates.

Due to these difficulties arising from our ignorance concerning the
cost and speed of future quantum hardware, I will simply consider a
quantum and a classical engine of same size (same number of gates) and
compare the number of clock cycles they need to find a desired
dataset. (Given my pessimism concerning cost and speed of quantum
hardware, this gives a big advantage to the quantum solution.) In this
setting it is clear that a quantum solution will be faster, the
question is only by how much.

\subsection{the architecture of my search engines}

The architecture of the best classical and quantum search engines I
can think of is essentially the same. The addressing system is a
binary tree, at each node of which there is a switch. At the top (the
root) there is a central processor and at the bottom (the leaves) we
have the $N$ datasets. As I consider (partially) parallel search
engines, I also place a processor at each node on some intermediate
level. These are identical processors which compute the search
criterion. Each of these processors sits atop $n$ datasets.

\begin{picture}(300,250)

\put(120,250){central processor} 

\thicklines
\put(150,245){\line(2,-1){80}} \put(150,245){\line(-2,-1){80}}

\put(70,205){\line(1,-1){40}} \put(70,205){\line(-1,-1){40}}
\put(230,205){\line(1,-1){40}} \put(230,205){\line(-1,-1){40}}
\put(30,155){P} \put(110,155){P} \put(190,155){P} \put(270,155){P} 

\put(-45,172){$N/n$}
\put(-45,160){search criterion}
\put(-45,148){processors}

\put(285,160){\quad each P:}
\put(285,148){space $p$, time $t$}

\put(30,150){\line(-2,-3){20}} \put(30,150){\line(2,-3){20}}
\put(110,150){\line(-2,-3){20}} \put(110,150){\line(2,-3){20}}
\put(190,150){\line(-2,-3){20}} \put(190,150){\line(2,-3){20}}
\put(270,150){\line(-2,-3){20}} \put(270,150){\line(2,-3){20}}

\put(10,120){\line(-1,-2){12}} \put(10,120){\line(1,-2){12}}
\put(50,120){\line(-1,-2){12}} \put(50,120){\line(1,-2){12}}
\put(90,120){\line(-1,-2){12}} \put(90,120){\line(1,-2){12}}
\put(130,120){\line(-1,-2){12}} \put(130,120){\line(1,-2){12}}
\put(170,120){\line(-1,-2){12}} \put(170,120){\line(1,-2){12}}
\put(210,120){\line(-1,-2){12}} \put(210,120){\line(1,-2){12}}
\put(250,120){\line(-1,-2){12}} \put(250,120){\line(1,-2){12}}
\put(290,120){\line(-1,-2){12}} \put(290,120){\line(1,-2){12}}

\put(233,92){$\underbrace{\hspace{74pt}}_{\mbox{$n$ datasets}}$}
\put(-10,72){$\underbrace{\hspace{320pt}}_{\mbox{$N$ datasets}}$}

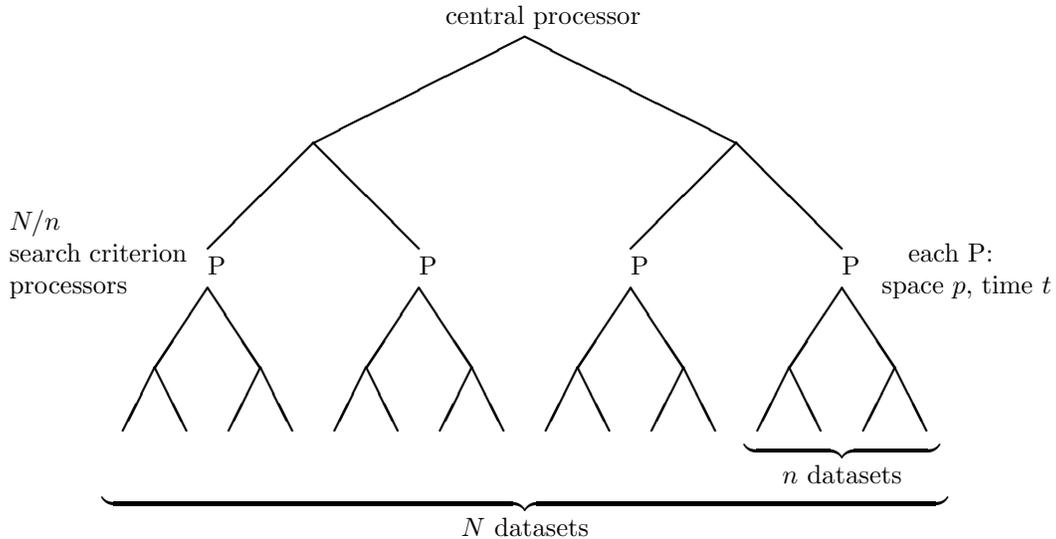
\put(-10,27){Figure 1: architecture of both, 
the quantum and the classical search engines}

\end{picture} \\

My considerations are not really affected if instead of a binary tree
we have a tree with some other small number of fan-outs at each
node. On the other hand things may change if the fan-out becomes
large, as I will discuss later.

\subsection{performance of the classical search engine}

The size (number of gates) of the classical search engine is ($S$ for
``space''):

\begin{equation}
S_c=\frac{N}{n}~ p +N \qquad .
\end{equation}
Again $N$ is the size of the database, $n$ the number of datasets per
processor and $p$ the size of a processor. Thus the first term is the
hardware in the processors and the second term is the addressing
system which clearly is about of the size of the database. Note that
here I assume that the switches at the nodes of the addressing system
have the same cost as the gates in the processors. This seems
reasonable as long as either could perform the other's task, as then
we simply use the cheaper of the two. This seems to be true as long as
the switches in the addressing tree have only few states (=small
fan-out).

The search engine works by having all processors in parallel search
through their $n$ datasets, each time computing the search criterion
in $t$ time steps. I neglect the (logarithmic) time it takes to
propagate the answer from a successful processor to the central
processor. I also neglect the time it takes a processor to retrieve
one of its datasets, assuming that this is less than the time it takes
to compute the search criterion. Then the search time is simply:

\begin{equation}
T_c=n t \qquad .
\end{equation}

\subsection{performance of the quantum search engine}

As stated above, I take the quantum search engine to be of the same
size as the classical one, thus I choose the same parameter $n$, so:

\begin{equation}
S_q=S_c=\frac{N}{n} p +N \qquad .
\end{equation}
Where I neglect the (probably small) factors by which a reversible
processor is larger and uses more time steps. 

Each (now quantum) search criterion processor now runs Grover's
algorithm on its $n$ datasets. The processor retrieves the contents of
the datasets in quantum parallelism and also computes the search
criterion in quantum parallelism. After roughly $\sqrt{n}$ such steps
it finds a dataset meeting our search criteria if there is one. The
propagation of this result to the central processor can now be done
classically. Thus the addressing system has to be quantum only below
the level of the processors, but that doesn't help much as that's of
course where most hardware is. The search time is now about:

\begin{equation}
T_q=\sqrt{n}~ t \qquad .
\end{equation}

\subsection{comparison}

We still have the free parameter $n$ which is equivalent to the size
and thus to the cost we want to invest. If we want to use as little
hardware as possible, then we don't use any parallelism and have only
a central processor, thus $n=N$. In this case we get the familiar
quantum speed-up of $\sqrt{N}$, even though we don't get a square root
of the number of classical steps $t N$ but only of the number $N$ of
datasets.

I don't think it is reasonable to use only as little hardware as
possible as with a relatively small additional investment we can get
big speed-ups. As we have already invested in an addressing system the
size of the database, I think it makes sense to invest a similar
amount into parallel processors for computing the search
criterion. Equal investment means:

\begin{equation}
\frac{N}{n}~ p = N  \qquad \Rightarrow \quad n=p \qquad .
\end{equation}
As $n$ can't be larger than $N$ we really have $n=min(p,N)$.
Now the quantum speed-up becomes:

\begin{equation}
T_c/T_q = min(\sqrt{p},\sqrt{N}) 
\end{equation}
Thus it depends on the complexity of the search criterion. For even
larger investment in hardware (smaller $n$) the quantum speed-up
becomes still smaller.

\subsection{(quasi-) analog addressing systems}

So far I have only considered tree like addressing systems. In
(classical) reality addressing systems often look different. Hard
disks are addressed in a quasi-analog manner where the read-write head
can be in a large number of positions. A second analog variable in
this system is time which allows us to retrieve the correct dataset by
simply reading at the right time. If such addressing systems (quantum
or classical) with many-state elements are much cheaper than tree like
ones, the comparison shifts in favor of the quantum solution.

\subsubsection{a quasi-analog quantum addressing system}

Imagine e.g. a 2- or even 3-dimensional optical storage medium (the
database) which can be addressed by a light beam (intersecting light
beams in the 3-dimensional case). A quantum addressing system might
now produce a photon in a superposition of many positions, thus
querying the database in quantum parallelism. This superposition might
be produced by mirrors being in a superposition of different classical
positions. After going through the optical medium the photon could
e.g. go back through the same set of mirrors.

The optical medium would shift the phase, change the polarization or
modify another property of the photon depending on the stored
data. Note that as in any reasonable setup the (of course classical)
database would act as an exterior field on the quantum information,
thus avoiding unwanted entanglement with it.

If inexpensive ways could be found to put a photon into a
position-superposition according to the content of a quantum-address
register, Grover's algorithm might really prove to be useful in
building search engines.

\section{Short summary}

I assume tree-like addressing systems and assume that quantum hardware
will be as cheap and fast as classical hardware. Then by arguing that
we should invest as much hardware into parallel processors as there is
anyways in the addressing system, I get that the speed-up of a quantum
search engine over a classical search engine would be
$min(\sqrt{p},\sqrt{N})$, where $N$ is the size of the database and
$p$ is the size of a processor needed to compute the search criterion
on the datasets.

\section{Conclusion}

Even under favorable assumptions about quantum hardware the speed-up
of a quantum search engine is limited not only by the size of the
database but also by the complexity of the search criterion.
Furthermore it is not clear (at least not to me) that there are any
databases of practical interest which can't be indexed intelligently
so that at least the vast majority of queries can be done very
quickly. Therefore I would rather not expect Grover-based database
search ever to be of much use. 

Still it is possible that there is a technological ``window of
opportunity'' for such solutions. By this I mean a time when certain
advanced technologies are available while others are still not. In the
long run it is not clear whether passive hardware (storage) will
continue to be much cheaper than active hardware (processors,
addressing systems). It seems that at this point the database search
problem would anyways go away.

I would like to thank Daniel Gottesman who has contributed to my
analysis through discussions. He is less pessimistic than me about the
usefulness of Grover's algorithm for database search.

\end{document}